\documentstyle[aps,epsf]{revtex}

\begin{document}
\title{Photonic band gaps and defect states induced by
excitations of Bose-Einstein condensates in optical lattices}

\author{Karl-Peter Marzlin and Weiping Zhang}
\address{
School of Mathematics, Physics,
Computing and Electronics, Macquarie University, Sydney, NSW 2109,
Australia
}
\maketitle
\begin{abstract}
We study the interaction of a Bose-Einstein condensate,
which is confined in an optical lattice, with
a largely detuned light field propagating through the condensate. 
If the condensate is in its
ground state it acts as a periodic dielectric
and gives rise to photonic band gaps at optical frequencies.
The band structure of the combined system of condensed lattice-atoms
and photons is studied by using the concept of polaritons.
If elementary excitations of the condensate are present, they will
produce defect states inside the photonic band gaps. 
The frequency of localized defect states
is calculated using the Koster-Slater model.
\end{abstract}
$ $ \\
\section{Introduction}
The achievement of Bose-Einstein condensation in magnetic traps
\cite{experimente} has induced a great interest in the properties
of quantum atomic gases and their manipulation by
atom optic techniques. Although the latter are usually used
for laser cooling in the formation process of a
Bose-Einstein condensate (BEC), confinement in an optical dipole-trap has been
demonstrated only recently \cite{ketterle98}.
The all-optical confinement of a BEC provides a great potential
for the manipulation and application of BEC. In particular,
it opens the new opportunity to study atomic BECs
in optical lattices. In recent years experimentalists have made
great efforts to create a BEC in optical lattices. Although there are
presently still some technical problems to achieve this goal the rich physics
of uncondensed ultracold atoms in optical
lattices \cite{deutsch96} and quasi-crystals \cite{grynberg97}
has attracted great interest for both experimentalists and theorists. 
Recently, several theoretical papers dealing with condensates in
optical potentials have been published \cite{molmer98,zoller98}

In this paper we focus on a new aspect of this subject: light
propagation through a coherent condensate confined in an optical lattice.
Since the ground state of the condensate in a lattice potential
is periodic, it will act as a
periodic dielectric for laser light propagating through it. Thus
it will give rise to photonic band gaps at optical frequencies.

The phenomenon of photonic band gaps is a natural consequence of the
periodicity of the condensate. In fact, it also should occur for
uncondensed ultracold atoms in optical lattices. However, in the case
of a condensed atomic lattice what is interesting is that, because of
the macroscopic occupation of the ground state, a proper 
description of photonic band gaps is given in terms of polaritons
(an entangled coherent system composed of
superpositions of photons and excited atoms).
Furthermore, elementary excitations may be present in the lattice BEC. 
In general these excitations are no
longer periodic and will cause distortions of the perfect periodic
structure of the condensed atomic lattice. An excitation-induced defect
in the atomic lattice in turn causes the occurence of defect states
inside photonic band gaps of light propagating through the BEC.
In this sense, elementary excitations have a close analogy to lattice
defects in solid state physics which cause defect states in the electronic
band structure.

The paper is organized as follows. In Sec.~II we will derive
the equations of motion. 
In Sec.~III we consider the case where only the lattice laser beams are
present and seek for a periodic solution to the coupled equations of motion
describing the ground-state BEC and the lattice laser beams. 
This solution shows that the optical potential
experiences no back-reaction from the condensate if the latter has
settled down into its ground state. 
In this sense, the lattice laser beams just effectively act as a
constant periodic potential for the BEC. In Sec.~IV we consider the
propagation of a weak probe laser beam through the ground-state BEC and
derive the form of the lowest photonic band
gap for this beam by using polariton modes.
To examine the behaviour of a probe laser beam
propagating through a weakly non-periodic BEC a theory of
defect states for photonic band gaps is developed in Sec.~V
which is applied in Sec.~VI
to the Koster-Slater model for a localized elementary excitation.
Sec.~VII concludes the paper.
\section{Equations of motion}
The system under consideration consists of interacting two-level atoms
which are coupled to the electromagnetic field. 
This coupling is described by using the electric-dipole
and rotating-wave approximation so that
the corresponding second quantized Hamiltonian is given by
\begin{equation} 
  H = H_A + H_{\mbox{{\scriptsize NL}}} +  H_{\mbox{{\scriptsize E.M.}}}
      + H_{\mbox{{\scriptsize int}}} \; ,
\label{hamil}\end{equation} 
where 
\begin{equation} H_A := \int d^3 x \sum_{i=e,g} \Psi_i^\dagger 
   \{ \frac{\vec{p}^2}{2M} +V(\vec{x}) +E_i \} \Psi_i
\end{equation} 
describes the atomic center-of-mass motion. $V(\vec{x})$ denotes an
external potential. $M$ represents the atomic mass, and $E_i$,
$i=e,g$ are the internal energy levels for ground-state and excited
atoms, respectively. The corresponding field operators $\Psi_g$ and
$\Psi_e$ fulfill the commutation relations $[\Psi_i(\vec{ x})^\dagger ,
\Psi_j(\vec{y})] = \delta_{ij} \delta(\vec{x}-\vec{y})$.
$H_{\mbox{{\scriptsize NL}}}$ is the nonlinear part of the atomic
Hamiltonian which describes two-body collisions. For a dilute Bose
gas it can be approximated by
\begin{equation} H_A := {1\over 2} \int d^3 x \sum_{i,j=e,g}
   g_{ij}  \Psi_i^\dagger \Psi_j^\dagger \Psi_j \Psi_i \; ,
\end{equation} 
where $g_{ij} := 4\pi \hbar^2 a_{\mbox{{\scriptsize sc}}}^{(ij)}/M$
are coupling constants and $ a_{\mbox{{\scriptsize sc}}}^{(ij)}$
denote the scattering lengths for scattering between atoms in the 
internal state $i$ and $j$.

For the description of the electromagnetic field we use the
representation in terms of positive end negative frequency
parts of the vector potential, $\vec{A}(\vec{ x}) = \vec{A}^{(+)}
(\vec{ x})+ \vec{A}^{(-)}(\vec{x})$. This representation will turn
out to be convenient for the adiabatic elimination of excited atoms.
The Hamiltonian for the free electromagnetic field then takes the simple
form
\begin{equation} H_{\mbox{{\scriptsize E.M.}}} = 
   2 \varepsilon_0 \int d^3 x \sum_{a,b=1}^3 A_a^{(-)} (\hat{\omega}^2)_{ab}
   A_b^{(+)} \; .
\end{equation}
The positive and negative frequency parts are related by
$(\vec{A}^{(+)})^\dagger =\vec{A}^{(-)}$ and fulfill the commutation
relation $[A_a^{(+)}(\vec{x}), A_b^{(-)}(\vec{y}) ] = 
(\hbar/2 \varepsilon_0) \hat{\omega}^{-1} 
\delta_{ab}^T(\vec{ x}-\vec{ y})$, where $\delta_{ab}^T(\vec{ x}-\vec{ y})$
is the transverse delta function.
The {\em frequency operator}
$\hat{\omega}$ is a pseudo-differential operator whose action
is defined in momentum space by
\begin{equation}
  [\hat{\omega} \vec{A}](\vec{x}) =
  (2\pi)^{-3/2} \int d^3 k e^{i \vec{k}\cdot \vec{x}} 
   c |\vec{k}| \vec{A}(\vec{k})\; .
\end{equation} 
The physical interpretation of the frequency operator is simple. It just
multiplies a photon mode with its frequency $\omega(\vec{k}) = c |\vec{k}|$.
A more compact representation of $\hat{\omega}$ in position space is
given by $\hat{\omega} = c\sqrt{-\hat{\Delta}} = c|-i\nabla|$, 
where $\hat{\Delta}$ denotes the Laplace operator.

Using the positive and negative frequency part of the vector potential
the electric dipole coupling between the atoms and the electromagnetic field
in rotating-wave approximation can be written as
\begin{equation} H_{\mbox{{\scriptsize int}}} = 
   i \int d^3 x  \{ \Psi_g^\dagger \Psi_e 
         (\vec{d}^* \cdot \hat{\omega} \vec{A}^{(-)} )
   -  \Psi_e^\dagger \Psi_g  (\vec{d}\cdot \hat{\omega} \vec{A}^{(+)}) \} \; .
\end{equation}

The Heisenberg equations of motion derived from the Hamiltonian
(\ref{hamil}) are given by
\begin{eqnarray} 
  i \hbar \dot{\Psi}_e &=& \left \{ \frac{\vec{p}^2}{2M} + V + E_e +
     \sum_{j=e,g} g_{ej} \Psi_j^\dagger \Psi_j  \right \} \Psi_e
     -i \Psi_g (\vec{d}\cdot \hat{\omega} \vec{A}^{(+)})
  \label{edgl} \\
  i\hbar \dot{\Psi}_g &=& \left \{ \frac{\vec{p}^2}{2M} + V + E_g +
     \sum_{j=e,g} g_{gj} \Psi_j^\dagger \Psi_j  \right \} \Psi_g
     +i \Psi_e (\vec{d}^* \cdot \hat{\omega} \vec{A}^{(-)})
  \label{gdgl} \\
  i \dot{A}^{(+)}_a(\vec{x}) &=& \hat{\omega}
  A_a^{(+)}(\vec{x}) + \frac{i}{2 \varepsilon_0} \int d^3 y
  \Psi_g^\dagger (\vec{y}) \Psi_e(\vec{y})\sum_{b=1}^3 d^*_b \delta_{ab}^T
  (\vec{x}-\vec{y}) \label{adgl}
\end{eqnarray} 
\section{Lattice laser beams and BEC in ground state: decoupling of
the fields}
To analyse the interaction between the atoms and the lattice laser
beams in absence of an external potential ($V(\vec{x})=0$)
we restrict us to the particular case
where the atomic field is composed of condensed atoms, i.e., a
Bose-Einstein condensate. This allows us to make further
substantial simplifications. As is well-known a condensate can
be described by assuming that all atoms are in the same quantum
state $\psi_g$. This amounts in replacing the field operator
$\Psi_g$ in Eq.~(\ref{gdgl}) by the c-number field $\psi_g$ which
then fulfills a nonlinear Schr\"odinger equation. In addition,
we assume that the photon fluctuations of the lattice laser beams
are small and therefore not important for our case. This allows us to
replace the operator $\vec{A}^{(+)}$ by a corresponding classical field
vector $\vec{A}^{(+)}_L$ of the lattice laser beams.

We consider the regime of coherent interaction where the
electromagnetic field is detuned far away from
the atomic resonance frequency $\omega_{\mbox{{\scriptsize res}}} :=
(E_e-E_g )/\hbar$. Specifically, we assume that the detuning 
$\Delta_L := \omega_L - \omega_{\mbox{{\scriptsize res}}}$ 
(with $\omega_L := c|\vec{k}_L|$) of the lattice
laser beams is negative (red detuning) and its absolute value
much larger than any other characeristic
frequency of our system so that we can adiabatically eliminate the 
excited atoms \cite{zhang94,lenz94,lewenstein94,marzlin98b}.
This amounts in replacing the field operator for excited atoms by
\begin{equation} 
  \Psi_e \approx \frac{-i}{\hbar \Delta_L} \psi_g  
  (\vec{d}\cdot \hat{\omega} \vec{A}^{(+)}_L) \; .
\label{adiab} \end{equation}
Inserting Eq.~(\ref{adiab}) into Eqs.~(\ref{gdgl}) and (\ref{adgl})
one easily finds (to first order in $1/\Delta_L $)
\begin{eqnarray} 
  i\hbar \dot{\psi}_g &=& \left \{ \frac{\vec{p}^2}{2M} + V + E_g +
      g_{gg} \psi_g^\dagger \psi_g  
  + \frac{1}{\hbar \Delta_L} (\vec{d}\cdot \hat{\omega} \vec{A}^{(-)}_L)
   (\vec{d}\cdot \hat{\omega} \vec{A}^{(+)}_L)
  \right \} \psi_g
  \label{gdgl2} \\
  i (\dot{A}^{(+)}_L(\vec{x}))_a &=& \hat{\omega}
  (A_L^{(+)}(\vec{x}))_a + \frac{1}{2 \varepsilon_0 \hbar \Delta_L} \int d^3 y
  |\psi_g (\vec{y})|^2 
  (\vec{d}\cdot \hat{\omega} \vec{A}_L^{(+)}(\vec{y}))
  \sum_{b=1}^3 d^*_b \delta_{ab}^T
  (\vec{x}-\vec{y}) \label{adgl2}
\end{eqnarray} 
Eqs.(\ref{gdgl2}) and (\ref{adgl2}) describe the coherent coupling of
a ground-state atomic field to the lattice laser beams.
The physics implicit in these equations is straightforward.
The laser beams induce an optical potential
for ground state atoms which is proportional to the light
intensity ($\propto (\hat{\omega}\vec{A}_L)^2$). The atoms in turn act
on the electromagnetic field like a dielectric, where the
index of refraction is determined by the density $|\psi_g|^2$ 
of ground-state atoms.

We are interested in how the condensate affects the lattice laser beams
and the corresponding back-reaction in the optical potential. 
For simplicity, we take the laser beam
to be parallel to the x-axis.
Further, we assume that the BEC has settled to its ground-state which,
because of the periodicity of the optical potential provided by the
lattice laser beams, is periodic. It is then convenient to decompose
the fields into a discrete Fourier series
\begin{eqnarray} 
  \psi_g(x) &=& \sum_{l\in {\bf Z}} \psi_l \exp [ i l \vec{k}_L\cdot
  \vec{x}] \nonumber \\
  \Omega^{(L)} (x) &:=& {1\over \hbar} \vec{d}\cdot \hat{\omega} 
  \vec{A}^{(+)}_L(x) =
  \sum_{l\in {\bf Z}} \Omega_l^{(L)} \exp [ i l \vec{k}_L\cdot
  \vec{x}] \; . \nonumber 
\end{eqnarray}  
We remark that $\vec{k}_L$ does {\em not} denote the wavevector of
the laser beams. It is defined by its relation to the spatial period 
$x_L$ of the optical lattice by $\vec{k}_L = \vec{e}_x 2\pi/x_L$.
This period $x_L$ differs in general slightly from the wavelength of
the laser beams outside the atomic medium \cite{fourwave}.
Transforming Eqs.~(\ref{gdgl2}) and (\ref{adgl2})
to momentum space we arrive at the following one-dimensional set
of equations,
\begin{eqnarray} 
  i \hbar \dot{\psi}_l &=& \frac{\hbar^2 k_L^2}{2M} l^2 \psi_l +
  \frac{g_{gg}}{(2\pi)^2} \sum_{m,n \in {\bf Z}} \psi^*_{m+n-l}
  \psi_m \psi_n  + \frac{\hbar}{(2\pi)^3 \hbar \Delta_L}
  \sum_{m,n \in {\bf Z}} \psi_{l+m-n} \Omega^{(L)*}_m \Omega^{(L)}_n
  \label{momg}\\
  i\hbar \dot{\Omega}^{(L)}_l &=& \hbar \omega_L |l| \Omega^{(L)}_l  + 
  \frac{\omega_L |l|\vec{d}_\perp^2
  }{2 (2\pi)^3 \varepsilon_0 \Delta_L } \sum_{m,n \in {\bf Z}}
   \Omega^{(L)}_m \psi^*_{m+n-l} \psi_n \; , \label{moma}
\end{eqnarray} 
where we have defined the transversal dipole moment of the atoms
$\vec{d}_\perp := \vec{d} - \vec{k}_L (\vec{d}\cdot \vec{k}_L)
/\vec{k}_L^2$.
We can now exploit the fact that the optical frequency $\omega_L$
is typically (very) much larger than any other frequency scale involved
in the system. This allows us to perform a rotating-wave approximation
by inserting $\tilde{\Omega}^{(L)}_l := \exp \{i \omega_L t |l|\} 
\Omega^{(L)}_l$
into Eqs.~(\ref{momg}) and (\ref{moma}) and neglecting all terms
which rotate at multiples of the frequency $\omega_L$. This procedure
results in the simplified equations
\begin{eqnarray} 
  i \hbar \dot{\psi}_l &=& \left \{ 
  \frac{\hbar^2 k_L^2}{2M} l^2 +
  \frac{\hbar}{(2\pi)^3 \Delta_L}
  \sum_{m \in {\bf Z}}  | \Omega^{(L)}_m|^2  \right \}\psi_l + 
  \frac{g_{gg}}{(2\pi)^2} \sum_{m,n \in {\bf Z}} \psi^*_{m+n-l}
  \psi_m \psi_n  
  \nonumber \\ & &
  + \frac{\hbar}{(2\pi)^3 \Delta_L}
  \sum_{m\in {\bf Z}} \psi_{l+2m} \Omega^{(L)*}_m \Omega^{(L)}_{-m}
  \label{mom2g}\\
  i \hbar \dot{\Omega}^{(L)}_l &=& \hbar \omega_L |l|\left \{ 1 + 
  \frac{\vec{d}_\perp^2}{2 (2\pi)^3 \varepsilon_0  \Delta_L } 
  \sum_{n \in {\bf Z}} | \psi_n|^2
  \right \}  \Omega^{(L)}_l  + 
  \frac{\vec{d}_\perp^2}{2 (2\pi)^3 \varepsilon_0  \Delta_L } 
  \omega_L |l| \Omega^{(L)}_{-l} \sum_{n \in {\bf Z}}
  \psi^*_{n-2l} \psi_n \; . \label{mom2a}
\end{eqnarray} 
These equations have some properties which allow to decouple
the system for some physically interesting cases. The most
important one is that the atoms do only couple counterpropagating
modes, i.e., $\Omega^{(L)}_l$ and $\Omega^{(L)}_{-l}$. 
This is a direct consequence
of energy conservation since a transition to any other mode would
require an amount of energy in the order of $\hbar \omega_L$ which cannot be
provided by the interaction with the condensate. In addition, it is not
difficult to see that both the mean density $\bar{\rho} :=
\sum_n |\psi_n|^2$  and the mean light intensity per mode,
$\bar{I}_l :=
|\Omega^{(L)}_l|^2 + |\Omega^{(L)}_{-l}|^2$, are conserved quantities, 
reflecting the conservation of the total number of atoms and 
the number of photons with energy $\hbar \omega_L |l|$, respectively.
Therefore, the first sums on the right-hand-side of Eqs.
(\ref{mom2g}) and (\ref{mom2a}) just produce a constant shift of
the energy levels.

We now consider a solution of the system 
of equations (\ref{mom2g}) and (\ref{mom2a}) which corresponds to
a standing-wave lattice laser beam
interacting with a BEC in its
ground state in the coherent regime. 
For a standing-wave lattice, we can make
the ansatz $\Omega^{(L)}_l = \Omega^{(L)}_{-l}$. 
In addition, in the ground state of the
BEC the time dependence of all coefficients $\psi_l$
is given by $\psi_l(t) = \exp[-i \mu t/\hbar] \psi_l(0)$, where
$\mu$ denotes the chemical potential and $\psi_l(0)$ 
can be chosen to be real.
It is then not difficult to show that $\Omega^{(L)}_l = \Omega^{(L)}_{-l}$
holds for all times (by differentiating both sides and comparing the
results). In addition, the expression $\Omega^{(L)}_l  \Omega^{(L)*}_{-l}$,
which describes the optical potential in Eq.~(\ref{mom2g}),
is time-independent, too.

An immediate consequence of this fact is that in the coherent regime
the optical potential 
created by the lattice laser beams for the BEC 
in Eq.~(\ref{mom2g})
is not altered by the condensate itself. Thus, the problem is
decoupled and the condensate behaves as if it were moving in a
given external periodic potential. 
In this sense the influence of the lattice laser beams on the BEC can be
replaced by an external potential $V(\vec{x})$ with periodicity
$\pi/k_L$.
\section{Polariton band theory for light interacting with a condensate}
In the preceding section we have shown how the combined system of a BEC
and lattice laser beams behaves in its ground state.
We are now interested in a different situation where a
running probe laser beam propagates through the ``lattice condensate''
(condensate plus lattice laser beams).
The behaviour of the probe laser can intuitively understood by
considering the BEC as a kind of dielectric for the probe laser beam.
Since the BEC is periodic the probe laser beam will effectively
propagate through a periodic dielectric. We thus expect it to show
the phenomenon of photonic band gaps \cite{yablonovitch91}.

To describe the interaction of the probe laser beam with the condensate
we will assume that the ground-state BEC changes little so that
$\psi_g$ enters as a given external field into the equations of motion
for excited atoms (\ref{edgl}) and for the probe laser beam
(\ref{adgl}). Solutions of these coupled equation describe polariton
modes, i.e., entangled superpositions of matter and light fields
\cite{shlyapnikov91,politzer91}. 
Thus, it is really ``polaritonic'' band gaps rather than photonic band gaps
that we are studying. However,
for sufficiently large detuning of the lattice laser beams the
entanglement is very small so that the result indeed can be considered as
photonic band gaps.

To find a suitable expression for the ground-state wavefunction $\psi_g$
we use the results of the preceding section, i.e.,
consider the case where the BEC is confined
by a periodic potential of the form
\begin{equation} 
V(\vec{x}) = -V_0 \cos (2 k_L x)
\end{equation} 
(we choose the factor of $2k_L$ since the potential created by an optical
lattice of wavevector $k_L$ would create such a potential
\cite{marzlin98b}). In addition, we consider a very weak probe beam
and neglect the four-wave mixing effect due to the
interference between
the probe laser and the lattice lasers \cite{fourwave}.
As a result, the ground state of the BEC can effectively be described
by the Gross-Pitaevskii equation
\begin{equation}
  \mu \psi_g = \left \{ \frac{\vec{p}^2}{2M} + V(\vec{x}) \right \}
  \psi_g + g_{gg} |\psi_g|^2 \psi_g\; ,
\label{gpe} \end{equation} 
where $\mu$ is the chemical potential.
In the experimentally
realized dilute Bose condensates the interaction energy of the
two-body collisions between atoms is usually large, so that one can
perform the Thomas-Fermi approximation by neglecting the kinetic energy.
This transforms Eq.~(\ref{gpe}) into a simple algebraic equation
whose solution is of the form  
$ |\psi_g(x)|^2 = \rho_0 + \rho_1 \cos (2 k_L x)$. 
This solution is valid for all $x$ if the optical potential is not too
strong, so that $\rho_1$ is smaller than $\rho_0$. If $\rho_1$
is not too close to $\rho_0$ we furthermore can simplify the wavefunction to
\begin{equation} 
  \psi_g(x) \approx \sqrt{\rho_0} + \frac{\rho_1}{2 \sqrt{\rho_0}}
  \cos (2 k_L x) 
\label{rho} \end{equation} 
(this time-independent expression is valid in a frame rotating at
frequency $E_g/\hbar$).
We remark that this expression produces qualitatively correct results
for the lowest photonic band even if the kinetic energy is not negligible
or $\rho_1\approx \rho_0$
because the corresponding correction essentially introduce higher 
coefficients in the Fourier
series of Eq.~(\ref{rho}). Since these higher coefficients do only couple
higher bands, they do not affect the results for the lowest band. 

Since $\psi_g$ does not depend on $y$ and $z$ it is advantageous to rescale
the wavefunctions as $\psi \rightarrow L_\perp \psi$, where
$L_\perp$ is the typical extension of the BEC in the $y$ and $z$
direction. This guarantees that the one-dimensional integral
$\int dx |\psi|^2$ is dimensionless and can be interpreted as a
particle number. In the actual calculations this rescaling
leads to the appearance of various factors of $L_\perp$.
$L_\perp$ will not enter the final results, though.

Introducing the (classical) field $\Omega^{(P)}(x) :=
\vec{d}\cdot \hat{\omega} \vec{A}^{(+)}_P(x)$ for the probe laser's
Rabi frequency Eqs.~(\ref{edgl}) and (\ref{adgl}) can be reduced to
the polariton equations of motions \cite{politzer91}
\begin{eqnarray} 
  i \hbar \dot{\psi}_e &=& \left \{ \frac{\vec{p}^2}{2M} + 
  \hbar \omega_{\mbox{{\scriptsize res}}}+
  {g_{eg}\over L_\perp^2} |\psi_g|^2  \right \} 
  \psi_e -i\hbar  \psi_g \Omega^{(P)}
  \label{edgl3} \\
  i \hbar \dot{\Omega}^{(P)} &=& \hbar \hat{\omega}
  \Omega^{(P)} + \frac{i |\vec{d}_\perp|^2}{2 \varepsilon_0 L_\perp^2} 
  \hat{\omega} \{ \psi_g^* \psi_e\} \; , \label{adgl3}
\end{eqnarray} 
where $\psi_g$ (divided by $L_\perp$) is given by Eq.~(\ref{rho}). 
Since the density of excited atoms should be very small we have neglected
two-body collisions between excited atoms ($g_{ee}=0$ in Eq.~(\ref{edgl})). 
Because $\psi_g$ only depends
on $x$, and since we consider the case that $\psi_e$ and $\Omega^{(P)}$
do not depend on $y$ and $z$ either, the transverse delta function of
Eq.~(\ref{adgl}) can be reduced to an ordinary delta function
(to prove this one can transform Eq.~(\ref{adgl}) to momentum space).

For later use it will be convenient to consider the solutions 
$\langle x | \phi \rangle := ( \psi_e(x), \Omega^{(P)}(x)\; )$ of
Eqs.~(\ref{edgl3}) and (\ref{adgl3}) as elements of a polariton
Hilbert space with conserved scalar product
\begin{equation}
  \langle \phi^\prime | \phi \rangle  := 
  \int dx \left \{ \psi_e^{\prime *} \psi_e + \frac{2\varepsilon_0 \hbar
  L_\perp^2}{
  |\vec{d}_\perp|^2} \Omega^{(P)\prime *} \hat{\omega}^{-1} \Omega^{(P)}
  \right \} \;.
\label{sprod} \end{equation} 
Physically the quantity $\langle \phi | \phi \rangle$ is related to
the number of excitations (number of excited atoms plus 
number of photons) in our system. It is a conserved quantity
because of the rotating-wave approximation made in Sec.~II.
Eqs.~(\ref{edgl3}) and (\ref{adgl3}) can be rewritten in the form
$i\hbar \partial_t |\phi \rangle = H_{\mbox{{\scriptsize pol}}}|\phi
\rangle $ with the polariton Hamiltonian
\begin{eqnarray} 
  H_{\mbox{{\scriptsize pol}}} & := & {1\over 2} ({\bf 1} + \sigma_3)
  \left \{ \frac{\vec{p}^2}{2M} + 
  \hbar \omega_{\mbox{{\scriptsize res}}} + 
  {g_{eg}\over L_\perp^2} |\psi_g|^2  \right \} 
  +   {1\over 2} ({\bf 1} - \sigma_3) \hbar \hat{\omega}
  -i\hbar  \psi_g \sigma_+
  + \frac{i |\vec{d}_\perp|^2}{2 \varepsilon_0 L_\perp^2} 
  \hat{\omega} \psi_g^* \sigma_-  \; ,
\end{eqnarray} 
where $\sigma_i$ are the Pauli matrices.
We remark that $ H_{\mbox{{\scriptsize pol}}}$ is Hermitean with respect to
the scalar product (\ref{sprod}), i.e., $\langle \phi^\prime | 
H_{\mbox{{\scriptsize pol}}} \phi \rangle  = \langle 
H_{\mbox{{\scriptsize pol}}} \phi^\prime | \phi$.

To derive the polariton band-structure we have to find the eigenvalues
of the operator $ H_{\mbox{{\scriptsize pol}}}$. Since $\psi_g$ is periodic
$H_{\mbox{{\scriptsize pol}}}$ commutes with the operator of discrete
translations of amount $\pi/k_L$ and thus has a common set of
eigenvectors with this operator. The eigenvectors $|\phi_{n,q} \rangle$
therefore can be characterized by two quantum numbers 
$n\in \{ 0, 1, 2, \cdots \} $ and $q \in [-k_L, k_L ] $ which denote the
band index and the quasi-momentum, respectively. The eigenvalues
of the discrete translation operator are given by $\exp [i q \pi/k_L]$
and belong to eigenvectors which are simply given by momentum eigenstates
with momentum $ \hbar k_m := \hbar (q + 2 m k_L) $ for integer $m$.
The eigenvalues $\hbar \omega_{n,q}$ of the Hamiltonian can be found
by expanding Eqs.~(\ref{edgl3}) and (\ref{adgl3}) in this basis
and searching for stationary solutions. The corresponding equations,
\begin{eqnarray} 
  \hbar \omega \psi_e(k_m) &=& \left \{ \frac{\hbar^2 k_m^2}{2M} +
  \hbar \omega_{\mbox{{\scriptsize res}}}
  \right \} \psi_e(k_m) + {g_{eg}\over L_\perp^2} \left \{ \rho_0 \psi_e(k_m)
  + {\rho_1 \over 2} [ \psi_e(k_{m+1}) + \psi_e(k_{m-1})]\right \}
  \nonumber \\ & &
  -i\hbar L_\perp \sqrt{\rho_0} \Omega^{(P)}(k_m) -i\hbar 
  L_\perp \frac{\rho_1}{4 \sqrt{
  \rho_0}} [  \Omega^{(P)}(k_{m+1}) + \Omega^{(P)}(k_{m-1}) ]
  \\
  \hbar \omega  \Omega^{(P)}(k_m) &=& \hbar c |k_m|  \Omega^{(P)}(k_m) +
  i \frac{|\vec{d}_\perp|^2 c |k_m|}{2 \varepsilon_0 L_\perp} \left \{
  \sqrt{\rho_0} \psi_e(k_m) + \frac{\rho_1}{4 \sqrt{\rho_0}} [
  \psi_e(k_{m+1}) + \psi_e(k_{m-1})] \right \} \; ,
\end{eqnarray} 
can easily be solved numerically. Fig.~1c) shows the resulting band structure
near the upper band edge of the lowest frequency band for a 
condensate of density $\rho_0 = 1.1 \rho_1 =
10^{14}$ cm$^{-3}$. In order to describe the limit of a photonic instead
of a polariton band-structure we have assumed a very large
detuning of $\Delta_L = 100$ GHz of the lattice. However,
the results given below do not change very much if a smaller detuning
is assumed. We furthermore have set
$|\vec{d}_\perp| \approx e a_0$, with $e$ being the electron's
charge and $a_0$ denoting Bohr's radius. The wavevector of the
lattice was taken to be $k_L = 10^{7}$ m$^{-1}$.

An excellent analytical approximation for the band structure can be made
by assuming that for $q\in (0,k_L)$ only the modes $\Omega^{(P)}(q)$,
$\Omega^{(P)}(q-2k_L)$, $\psi_e(q)$, and $\psi_e(q-2k_L)$ are important.
The problem is then reduced to finding the eigenvalues of a $4\times 4$
matrix. For $q=k_L$ the eigenvalues have a simple form and allow to derive
the following expression for the band gap $\Delta \omega$ separating the
two lowest energy bands,
\begin{equation} 
  \Delta \omega = s_+ - s_-  \; ,
\end{equation} 
where we have defined the frequencies $\nu_i := |\vec{d}_\perp|^2 \rho_i
/(2 \hbar \varepsilon_0)$
and introduced the abbreviations 
$s_{\pm} := \sqrt{(\Delta_L / 2 )^2  +
   \omega_{\mbox{{\scriptsize res}}} (\sqrt{\nu_0}\pm\sqrt{\nu_1}/4)^2}$.
For a large detuning $|\Delta_L|$, i.e., in
the limit of a photonic band gap, this expression simplifies to
$\Delta \omega = \omega_0 \nu_1/|\Delta_L|$. For the numerical
values given above the band gap takes the value $\Delta \omega \approx$
40 GHz.

For $q\neq k_L$ the band structure is given by a rather complicated
expression. We therefore have
further simplified the analytical result by fitting it to a square root
\cite{fitting}. The lowest polariton band then takes the form
\begin{equation} 
  \omega_{0,q} \approx \omega_{0,\mbox{{\scriptsize max}}} + \bar{\nu}
  -\sqrt{c^2(|q|-k_L)^2 + \bar{\nu}^2} \; ,
\label{band} \end{equation} 
where
\begin{equation} 
   \omega_{0,\mbox{{\scriptsize max}}} =  \omega_{\mbox{{\scriptsize res}}}
   - \frac{|\Delta_L |}{2} - s_+ 
\end{equation} 
denotes the upper edge of the lowest frequency band, and
\begin{equation} 
  \bar{\nu} := \frac{\Delta \omega}{ (\omega_{0,\mbox{{\scriptsize max}}}
  -\omega_{\mbox{{\scriptsize res}}})^2} s_+(s_+ + s_-)
\end{equation} 
determines the curvature of the band. In the limit of a large 
detuning this simplifies to $\bar{\nu} \approx \Delta \omega /2$.
\section{Theory of localized defects} 
In the previous section we have studied polariton band gaps of
light generated by the lattice condensate in its ground state.
However, in general the BEC might be in a state corresponding
to a (coherent) elementary excitation which usually are not periodic.
Thus, we expect defects in the lattice condensate.
As is well-known from solid state theory a defect or an impurity in an
otherwise periodic potential can lead to defect states, i.e., 
states whose energy eigenvalue lies inside the gap between two 
energy bands (see, e.g., Ref.~\cite{callaway91}).  
In the system under consideration this phenomenon could be exploited to
acquire knowledge about non-periodic elementary excitations of the
Bose condensate by observing light propagation
through the BEC. In addition, defect theory can also be applied
to study the back-reaction of an excited BEC on the optical potential
confining it.

The existence of defect states for photonic band gaps has been
examined in the microwave regime 
for ordinary dielectric materials 
\cite{yablonovitch91,defectpaper}.
The method of calculation that we adopt is closely related to
the Green's function approach of Ref.~\cite{callaway91}.

Specifically we consider the situation that the condensate's wavefunction
is given by $\psi_G(x) = \psi_g(x) +
\delta \psi_g(x)$, where $\psi_g$ (divided by $L_\perp$)
is given by Eq.~(\ref{rho})  and $\delta \psi_g$
describes a coherent
elementary excitation of the condensate which we assume to be
localized in the x-direction. This
allows us to estimate the resulting energy eigenstates by adapting
the Koster-Slater model \cite{koster54} to the case of polariton
band gaps.

Our aim is to find solutions of the equation
\begin{equation} 
   \hbar \omega_{\mbox{{\scriptsize defect}}}
   |\phi \rangle = (H_{\mbox{{\scriptsize pol}}}
   + H_{\mbox{{\scriptsize defect}}}) |\phi \rangle\; ,  
\label{defectdgl} \end{equation} 
where
\begin{equation} 
  H_{\mbox{{\scriptsize defect}}} :=  -i\hbar \delta \psi_g \sigma_+
  + \frac{i |\vec{d}_\perp|^2}{2 \varepsilon_0 L_\perp^2} 
  \hat{\omega} \delta \psi_g^* \sigma_- 
\end{equation} 
describes the influence of the elementary excitation $\delta \psi_g$
and $\omega_{\mbox{{\scriptsize defect}}}$ is the defect eigenfrequency.

To find the solutions of Eq.~(\ref{defectdgl}) we expand
$|\phi \rangle $ in terms of Wannier functions,
\begin{equation} 
  |W_{n,\nu} \rangle  := {1\over \sqrt{2k_L}} \int_{-k_L}^{k_L}
  dq e^{-\pi i \nu q/k_L} |\phi_{n,q} \rangle \; ,
\end{equation} 
where $\nu$ is an integer number.
The functions $W_{n,\nu}(x)$ are localized around the lattice point
$\pi \nu/k_L$ which makes them a convenient tool to study
localized defects. Inserting $|\phi \rangle = \sum_{n,\nu}
\phi_{n,\nu} |W_{n,\nu} \rangle $ into Eq.~(\ref{defectdgl})
one easily deduces the equation
\begin{equation} 
  \omega_{\mbox{{\scriptsize defect}}}
  \phi_{n,\nu} = \sum_{\mu} \omega_{n,\nu-\mu} \phi_{n,\mu}
  + \sum_{m,\mu} \langle W_{n,\nu} |  
   H_{\mbox{{\scriptsize defect}}}/\hbar |W_{m,\mu} \rangle \phi_{m,\mu}\; , 
\label{w1} \end{equation} 
with
\begin{equation} 
  \omega_{n,\nu-\mu} := {1\over 2 k_L} \int_{-k_L}^{k_L} dq
  e^{\pi i(\nu-\mu)/k_L} \omega_{n,q} \; .
\end{equation} 
Since the Wannier basis is countable Eq.~(\ref{w1}) can be
interpreted as a matrix equation. In particular, one now can exploit
the fact that both the Wannier functions and $H_{\mbox{{\scriptsize defect}}}$ 
are localized. This implies that, at least approximately, only
a finite number, say an $N\times N$ submatrix, of the matrix elements
$ \langle W_{n,\nu} |H_{\mbox{{\scriptsize defect}}} |W_{m,\mu} \rangle$
is nonvanishing. 

Introducing the Green's function $G := ( \omega_{\mbox{{\scriptsize defect}}}
 {\bf 1} - H_{\mbox{{\scriptsize pol}}}/\hbar )^{-1}$ 
whose matrix elements are given by
\begin{equation} 
  \langle W_{n,\nu} | G | W_{m,\mu} \rangle = \delta_{n,m} {1\over 2k_L}
  \int_{-k_L}^{k_L} dq \frac{e^{\pi i (\nu-\mu)/k_L}}{
  \omega_{\mbox{{\scriptsize defect}}} -\omega_{n,q}} 
  =: \delta_{n,m} G_{n, \nu-\mu}
\label{gmatelem} \end{equation} 
the eigenvalue problem can be reduced to
\begin{equation} 
  \phi_{n,\nu} = \sum_{m,\mu,\lambda} G_{n, \nu-\mu}
   \langle W_{n,\mu} |  
   H_{\mbox{{\scriptsize defect}}}/\hbar |W_{m,\lambda} \rangle 
   \phi_{m,\lambda}   \; .
\label{evbed} \end{equation} 
As is well-known in solid state theory \cite{callaway91}
it is sufficient to consider only the eigenvalue problem of the
$N\times N$ subspace where the matrix elements of
$ H_{\mbox{{\scriptsize defect}}}$ are non-zero to derive the frequencies
of the defect states.
\section{Koster-Slater model for photonic band gaps}
Having derived the matrix eigenvalue equation (\ref{evbed}) for
the determination of the defect frequency it is straightforward
to apply the Koster-Slater model \cite{koster54} to the problem
at hand. In this model the assumption is made that both the Wannier
functions and the perturbation $U_{\mbox{{\scriptsize np}}}$
are localized in such a way that only one matrix element of the perturbation
is nonzero,
\begin{equation} 
  \langle W_{n,\nu} | H_{\mbox{{\scriptsize defect}}} |W_{m,\mu} \rangle
  = U_0 \delta_{n,0} \delta_{m,0} \delta_{\nu,0} \delta_{\mu,0} \; .
\label{ksm} \end{equation} 
This model is not valid if $ \delta \psi_g$ is 
too strongly localized (i.e., on a scale much smaller than the lattice
spacing $\pi/k_L$) \cite{stoneham75}, but should produce qualitative
estimates of defect frequencies for moderately localized perturbations.

Inserting Eq.~(\ref{ksm}) into the eigenvalue equation (\ref{evbed})
and using Eq.~(\ref{gmatelem}) to evaluate the only relevant matrix
element of $G$, $\langle W_{0,0}| G | W_{0,0} \rangle $, we find
that the defect frequency $\omega_{\mbox{{\scriptsize defect}}}$ 
has to fulfill the condition
\begin{equation} 
  1 = {U_0\over \hbar} \frac{1}{2k_L} \int_{-k_L}^{k_L}
  \frac{dq}{\omega_{\mbox{{\scriptsize defect}}} -\omega_{0,q}} \; .
\label{ksbed} \end{equation}  
The integral can be calculated exactly for the photonic band
of Eq.~(\ref{band}), but since the resulting expression is somewhat
complicated it is more instructive to use the following
approximation which is valid if the defect frequency 
$\omega_{\mbox{{\scriptsize defect}}}$ is close
to the upper edge $\omega_{0,\mbox{{\scriptsize max}}}$
of the lowest frequency band,
\begin{equation} 
  \frac{1}{2k_L} \int_{-k_L}^{k_L} \frac{dq}{
  \omega_{\mbox{{\scriptsize defect}}} -\omega_{0,q}}
  \approx \frac{\pi}{\omega_L} \sqrt{\frac{\bar{\nu}}{2(
  \omega_{\mbox{{\scriptsize defect}}}-
  \omega_{0,\mbox{{\scriptsize max}}} ) }} \; .
\end{equation} 
Inserting this into Eq.~(\ref{ksbed}) we find for the frequency of the
defect state the expression
\begin{equation} 
  \omega_{\mbox{{\scriptsize defect}}} - 
  \omega_{0,\mbox{{\scriptsize max}}} \approx
  \frac{\pi^2}{2} \frac{\bar{\nu}}{\omega_L^2} \frac{U_0^2}{\hbar^2} \; .
\label{deffreq} \end{equation} 
It is of interest to know how large this frequency difference is for realistic
systems. To achieve this we first have to estimate the value of
$U_0 = \langle W_{0,0} |  H_{\mbox{{\scriptsize defect}}} | W_{0,0} \rangle
= i\hbar \int dx \{ W_\Omega^* W_e \delta \psi_g^* - W_e^* W_\Omega
\delta \psi_g \}$ with $\langle x|W_{0,0} \rangle = (W_e(x), W_\Omega(x))$.
A rough estimate for this integral can be made by setting both
$\delta \psi_g$ and the Wannier function to be constant over 
one wavelength $\lambda_L$ and to be zero outside this range.
The normalization condition $ \langle W_{0,0}|W_{0,0} \rangle =1$
then leads approximately to 
$1= \lambda_L \{ |W_e|^2 + |W_\Omega|^2/(\omega_L |d_\perp|^2
/2\hbar \varepsilon_0 L_\perp^2) \}$. Using this condition $U_0$ takes
its maximal value for $W_e = 1/(\sqrt{2}L_\perp)$,
\begin{equation} 
  U_0 \approx 2 \hbar \delta \psi_g \sqrt{\omega_L \frac{ |d_\perp|^2}{
  2 \hbar \varepsilon_0 L_\perp^2}} \; . 
\end{equation} 
Assuming a defect amplitude of  
$\delta \psi_g = \epsilon \sqrt{\rho_0} L_\perp$  over the
extent of $W_{0,0}(x)$, where $\epsilon$ is small compared to one,
we can derive an estimate for the defect frequency (\ref{deffreq}) of
\begin{equation} 
  \omega_{\mbox{{\scriptsize defect}}} - 
  \omega_{0,\mbox{{\scriptsize max}}} \approx
  2 \pi^2 \epsilon^2 \frac{\bar{\nu}}{\omega_L} \nu_0 \; .
\end{equation} 
Using the same numbers as in Sec.~IV (that is, $\rho_i = 10^{14}$ cm$^{-3}$
and $\Delta_L =10^{11}$ s$^{-1}$) as well as $\epsilon =0.3$ this
frequency difference can be shown to be of the order of 100 Hz. 
Though this number is too small to be measurable it should be pointed out
that it applies only to a defect corresponding to an elementary
excitation over one wavelength. A different type of defect can
produce a much different result. For example, if we do not consider
a weak elementary excitation but a strong localized excitation
we can estimate its effect but assuming a larger value for $\epsilon$.
A threefold increase of the local density ($\epsilon = 3$) would lead
to a defect frequency $\omega_{\mbox{{\scriptsize defect}}} - 
\omega_{0,\mbox{{\scriptsize max}}}$
of about 10 KHz, for instance.
\section{Conclusions}
In this paper we have analysed the interaction of a lattice (or
``crystalized'')
Bose-Einstein condensate with largely detuned laser beams. We have
derived a periodic solution of the coupled equations of
motion, corresponding to a free BEC and a standing lattice laser beam.
We found that, if the condensate is in its ground state,
these equations decouple and the effect of the lattice laser beam on the
condensate is not affected by the condensate itself (no back-reaction).
In this situation it is thus equivalent to consider a condensate in some
external periodic potential with the same periodicity.
 
Building on this result we then assumed that the condensate is confined
by an external periodic potential. Since
the condensate's ground state is then periodic, too, it forms a kind
of periodic dielectric. A probe laser beam propagating through this dielectric
will then experience the formation of photonic band gaps. We have
analysed this situation using the concept of polaritons, i.e., 
entangled superpositions of excited atoms and photons.

If the condensate is not in its ground state but in a state corresponding
to a localized elementary excitation the periodicity of the system is 
perturbed. This leads to the formation of defect states inside
a polariton band gap.

{\bf Acknowledgement}: We thank Michael Steel for helpful discussions.
 This work has been supported by the Australian Research Council.


\newpage
\begin{figure}[t]
\epsfxsize=6cm
\hspace{3cm}
\epsffile{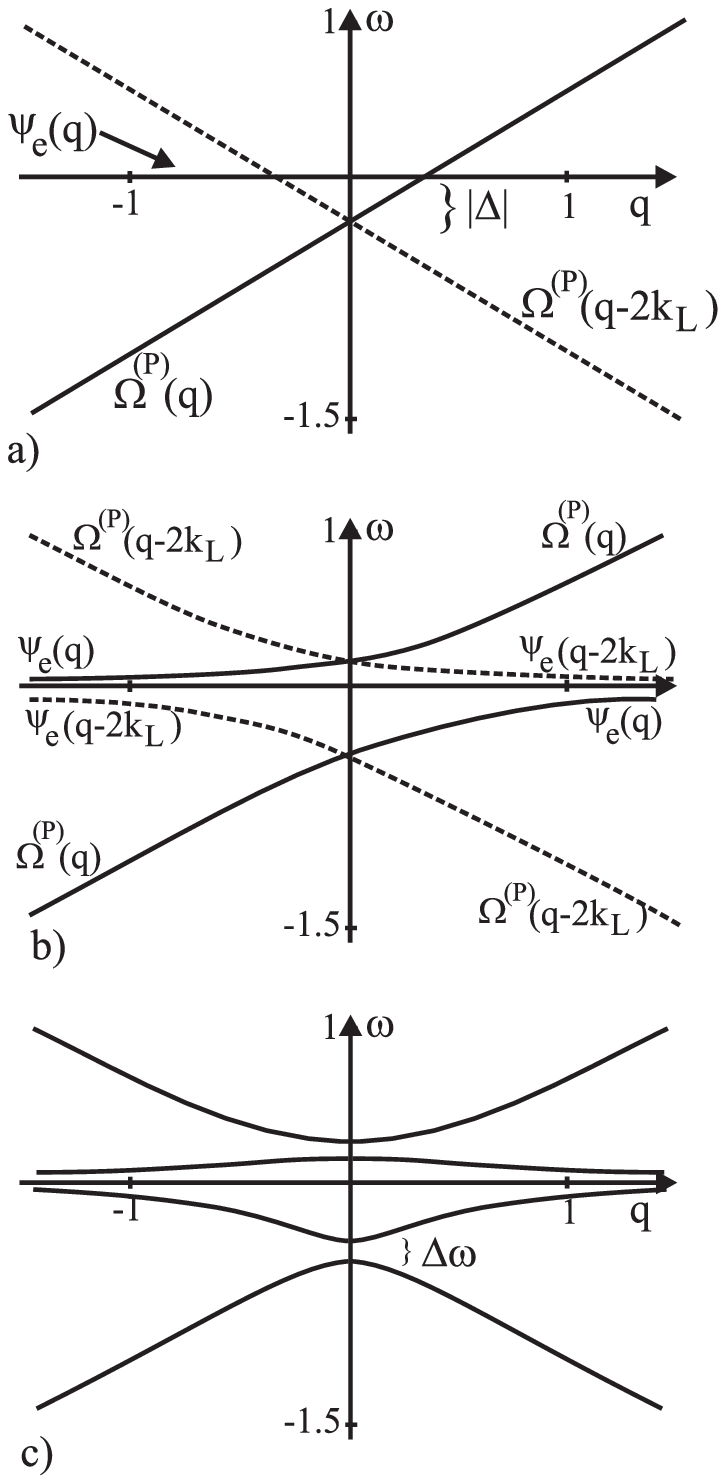}
\caption{Band structure for polaritons near the band edge $k_L$.
Displayed is $10^4 (\omega-\omega_{\mbox{{\scriptsize res}}})/
\omega_{\mbox{{\scriptsize res}}}$ versus $10^4 (q-k_L)/k_L$.
Fig.~a) shows the free dispersion relation for the relevant field components
in absence of any interaction. For a homogeneous BEC, $\psi_g = \sqrt{\rho_0}$,
a mixing of excited atoms and photons leads to the formation of
two separate avoided crossings (solid and dashed lines in b)).
The labels indicate the asymptotic form of the polariton modes, where
they correspond to either excited atoms ($\psi_e$) or photons
($\Omega^{(P)}$). For a periodic ground-state BEC the two avoided
crossings are combined to form band gaps (Fig.~c).}
\end{figure}

$ $

$ $


\begin{thebibliography}{99}
\bibitem{experimente} M. Anderson, J.R. Ensher, M.R. Matthews, 
	C.E. Wieman, and E.A. Cornell, Science {\bf 269}, 198 (1995);
	C.C. Bradley, C.A. Sackett, J.J. Tollet and R. Hulet, 
	Phys.~Rev.~Lett.~{\bf 75}, 1687 (1995);
	M.-O. Mewes, M.R. Andrews, N.J. van Druten, D.M. Kurn, 
	D.S. Durfee, C.G. Townsend and W. Ketterle, 
	Phys.~Rev.~Lett. {\bf 77}, 416 (1996);
	D.S. Jin, J.R. Ensher, M.R. Matthews, C.E. Wieman, 
	and E.A. Cornell, Phys.~Rev.~Lett.~{\bf 77}, 420 (1996).
\bibitem{ketterle98} D.M. Stamper-Kurn, 
        M.R. Andrews, A.P. Chikkatur,
	S. Inouye, H.-J. Miesner, J. Stenger, and W. Ketterle,
	Phys.~Rev.~Lett.~{\bf 80}, 2027 (1998).
\bibitem{deutsch96} P.S. Jessen and I.H. Deutsch, {\em Optical lattices}, in
	{\em Advances in atomic, molecular, and optical Physics}
	{\bf 37}, 95 (1996), edited by B. Bederson and H. Walther,
	Cambridge 1996.
\bibitem{grynberg97} L. Guidoni, 
        C. Trich\'e, P. Verkerk, and G. Grynberg,
	Phys.~Rev.~Lett.~{\bf 79}, 3363 (1997).
\bibitem{molmer98} K. Berg-S{\o}rensen and K. M{\o}lmer, to appear in
	Phys. Rev. A.
\bibitem{zoller98} D. Jaksch, C. Bruder, J.I. Cirac, C.W. Gardiner, 
	and P. Zoller, preprint cond-mat/9805037.
\bibitem{zhang94} W. Zhang and D.F. Walls, Phys. Rev. A {\bf 49},
	3799 (1994).
\bibitem{lenz94} G. Lenz, P. Meystre, and E.M. Wright, 
	Phys. Rev. A {\bf 50}, 1681 (1994).
\bibitem{lewenstein94} M. Lewenstein, L. You, J. Cooper, 
	and K. Burnett, Phys. Rev. A {\bf 50}, 2207 (1994).
\bibitem{marzlin98b} K.-P. Marzlin und W. Zhang,
	 Phys. Rev. A. {\bf 57}, p. 4761 (1998).
\bibitem{fourwave} I.H. Deutsch, R.J.C. Spreeuw, S.L. Rolston, and
	W.D. Phillips, Phys.~Rev.~A {\bf 52}, 1394 (1995).
\bibitem{yablonovitch91} E. Yablonovitch, T.J. Gmitter, R.D. Meade, 
	A.M. Rappe, K.D. Brommer, and J.D. Joannopoulos,
	 Phys.~Rev.~Lett.~{\bf 67}, 3380 (1991).
\bibitem{shlyapnikov91} B.V. Svistunov and G.V. Shlyapnikov,
	Sov. Phys. JETP {\bf 71}, 71 (1990).
\bibitem{politzer91} H.D. Politzer, Phys. Rev. A {\bf 43}, 6444 (1991).
\bibitem{fitting} More specifically, we assumed that the lowest frequency
	band $\omega_{0,q}$ has the form 
        $\omega_{0,q} \approx f(q) := A - \sqrt{c^2(|q|-k_L)^2 +B^2}$. 
        The parameters $A$ and $B$ are determined by comparing
        $f$ and $d^2 f/dq^2$ at $q=k_L$ with 
	$\omega_{0,q}$ and $d^2 \omega_{0,q}/dq^2$ found in the 4-level
	approximation. The latter can be found by taking the derivative
	of the characteristic polynomial of the $4\times 4$ matrix with
	respect to $q$ at $q=k_L$. 
\bibitem{callaway91} J. Callaway, {\em Quantum theory of the solid state},
	2nd edition, Academic Press, Boston 1991.
\bibitem{defectpaper} R.D. Meade, K.D. Brommer, A.M. Rappe, and 
	J.D. Joannopoulos, Phys.~Rev.~B {\bf 44}, 13772 (1991);
	K.M. Leung, J.~Opt.~Soc.~Am.~B {\bf 10}, 303 (1993);
	D.R. Smith, R. Dalichaouch, N. Kroll, S. Schultz,
 	S.L. McCall, and P.M. Platzman, 
	J.~Opt.~Soc.~Am.~B {\bf 10}, 314 (1993);
	H.G. Algul, M. Khazhinsky, A.R. McGurn, and J. Kapenga,
	 J.~Phys.: Condens. Matter {\bf 7}, 447 (1995);
	N.-H. Liu, Phys.~Rev.~B {\bf 55}, 4097 (1997).
\bibitem{koster54} G.F. Koster and J.C. Slater, Phys.~Rev.~{\bf 96},
	1208 (1954).
\bibitem{stoneham75} A.M. Stoneham, {\em Theory of defects in solids},
	Clarendon Press, Oxford 1975.
\end{thebibliography}
\end{document}